\documentclass[prb,twocolumn,a4paper,groupedaddress,showpacs,floatfix]{revtex4}
\usepackage{graphicx,amsmath,amssymb,amsfonts,dsfont,subfigure}

\usepackage{hyperref}
\usepackage{color}

\begin{document}

\title{Exciton transport enhancement across quantum Su-Schrieffer-Heeger lattices with quartic nonlinearity}

\author{J. J. Mendoza-Arenas$^{1}$, D. F. Rojas-Gamboa$^{1}$, M. B. Plenio$^2$ and J. Prior$^{3,4}$}

\affiliation{${}^1$Departamento de F\'{i}sica, Universidad de los Andes, A.A. 4976, Bogot\'a D. C., Colombia}
\affiliation{${}^2$Institut f\"ur Theoretische Physik \& IQST, Albert-Einstein Allee 11, Universit\"at Ulm, D-89081 Ulm, Germany}
\affiliation{${}^3$Area de F\'isica Aplicada, Universidad Polit\'ecnica de Cartagena, Cartagena 30202, Spain}
\affiliation{${}^4$Instituto Carlos I de F\'isica Te\'orica y Computacional, Universidad de Granada, Granada 18071, Spain}

\date{\today}

\begin{abstract}
In the present work we discuss the propagation of excitons across a one-dimensional Su-Schrieffer-Heeger lattice, which possesses both harmonic oscillations and weak quartic anharmonicities. When quantizing these vibrational degrees of freedom we identify several phonon-conserving non-linearities, each one with a different impact on the excitonic transport. Our analysis identifies a dominant non-linear correction to the phonon hopping which leads to a strong enhancement of exciton conduction compared to a purely linear vibrational dynamics. Thus quartic lattice non-linearities can be exploited to induce transitions from localized to delocalized transport, even for very weak amplitudes.
\end{abstract}

\pacs{05.60.Gg, 05.10.-a, 63.20.Ry, 68.65.-k}

\maketitle

\section{Introduction}

The dynamics of interacting many-particle systems is highly non trivial and continues to present unexpected and surprising features. The importance of the interaction of environmental phonons with electrons, the role of an effective electron cloud coming from multiple simultaneous electron transitions, and the intrinsic dynamics and relevance of a structured, evolving environment remains to be elucidated in detail. A standard model for open system dynamics considers such a system interacting with a discrete or continuous set of harmonic oscillators. Simple treatments trace out the environment, an approach that requires a number of approximations for its validity, and usually describe the dynamics by Markovian master equations~\cite{breuer}. On the other hand, more advanced schemes take full account of the environmental dynamics~\cite{TEDOPA, QUAPI, HEOM}, but are numerically more demanding and applicable to harmonic environments only. Hence, most studies have adopted the first approach, which results in Markovian decoherence~\cite{plenio2008njp,mohseni_jcp2008,chin2010njp,semiao2010njp,sinayskiy_prl2012,cai2013prl,we,we2,we3,santos2016pre,Levi2016prl,Fischer2016prl,Marko:2017ann,wolff:2018}, or consider dynamical properties for lattice harmonicity~\cite{chung2007prb,vidmar2011prb,golez2012prl,dorfner2015prb,brockt2015prb,hashimoto2017prb,brockt2017prb,tozer2012jpca,mannouch2018jcp}. Taking into account specific inter-atomic potential energy terms beyond this harmonic approximation remains a challenge.  

However, anharmonicity in quantum systems is ubiquitous. Firstly, it is crucial for explaining phenomena such as the high-temperature specific heat and thermal expansion of solids~\cite{ashcroft}, ferroeletricity~\cite{bianco2017prb,poojitha2019prm}, stabilization of crystal structures~\cite{errea2011prl}, and superconductivity of particular compounds~\cite{errea2013prl}. In addition, lattice non-linearities are critical for efficiently manipulating ordered phases in correlated systems, by means of mechanical~\cite{leroux2015prb} or optical~\cite{mankowski2016rep} protocols. Furthermore, they are known to play a key role in a wide variety of quantum transport processes as photon-assisted electronic conduction in nanostructures~\cite{aguado2004pr}, heat flow~\cite{saito2007prl}, vibrational energy transfer~\cite{leitner2001prb}, and metal-insulator transitions in complex materials~\cite{budai2014nat}.

Even though these efforts have evidenced the importance of anharmonicity for dynamical properties of quantum systems, an understanding of the effect of the several non-linear processes it induces is still lacking. Motivated by this gap, in the present work we carry out a systematic study of the impact of an anharmonic potential on the dynamics of a quantum system with an underlying oscillating structure. In particular we simulate the propagation of excitons across a one-dimensional vibrating lattice of ions, described by the archetypical Su-Schrieffer-Heeger (SSH) model~\cite{su1988rmp} with a quartic anharmonicity~\cite{saito2007prl,errea2011prl,errea2013prl,freericks2000prb,voulgarakis2000prb,Iubini2015njp,savelev2018prb}. We first determine the different types of particle-conserving non-linearities that emerge from the quantization of the vibrational degrees of freedom due to the quartic potential. Performing matrix product calculations, we then elucidate the influence of each term on the exciton dynamics. We observe that while some non-linearities tend to impede transport, others increase it, and that the overall effect is dominated by a phonon hopping renormalization. Our main conclusion is that it is possible to strongly enhance exciton transport even with a very weak anharmonic lattice potential.


The article is organized as follows. In Sec.~\ref{model_section} we discuss the model to be considered, where we show that different types of non-linear processes emerge when quantizing the anharmonic lattice potential. In Sec.~\ref{linear_section} we analyze the dynamics of an initial exciton-phonon excitation in the linear regime. Afterwards we discuss how the different non-linear mechanisms affect exciton dynamics, as presented in Sec.~\ref{nonlinear_section}, and see the effect of the total nonlinearity. Finally we show our conclusions in Sec.~\ref{conclu}.  

\section{Model} \label{model_section}
In the present work, we study the propagation of excitons through a non-linear one-dimensional oscillating lattice of ions. We now discuss how we model, simulate and characterize such a dynamical process when the ionic vibrational degrees of freedom are quantized by phonons.

\subsection{Exciton lattice with phonon nonlinearity}
The system under study is described by the sum of a Hamiltonian of non-interacting particles (in this case, excitons) propagating in a one-dimensional lattice, the Hamiltonian of the underlying ions, and the coupling between both. The total Hamiltonian is given by~\cite{Iubini2015njp}
\begin{equation}
H=H_{\text{ex}}+H_{\text{lat}}+H_{\text{in}}.
\end{equation}
Here $H_{\text{ex}}$ is the exciton Hamiltonian,
\begin{equation}
H_{\text{ex}}=J\sum_{j=1}^{N-1}\left(c_{j}^{\dagger}c_{j+1}+\text{h.c.}\right),
\end{equation}
with $c_{j}^{\dagger}$ ($c_j$) the creation (annihilation) operator of an exciton at site $j$, $J$ their hopping rate between nearest neighbors, which we take as $J=1$ to set the energy scale, and $N$ the number of sites; we also take $\hbar=1$ throughout our work. In addition, we consider that the lattice Hamiltonian $H_{\text{lat}}$ describing the dynamics of the underling  ions incorporates both linear and quartic non-linear terms, namely
\begin{equation} \label{lat_classical}
H_{\text{lat}}=\sum_{j=0}^{N+1}\frac{p_j^2}{2m}+\frac{\alpha}{2}\sum_{j=0}^{N}\left(u_{j+1}-u_{j}\right)^2+\frac{\lambda}{4}\sum_{j=0}^{N}\left(u_{j+1}-u_{j}\right)^4.
\end{equation}
Here $m$ is the ionic mass, $p_j$ the momentum of the ion at site $j$, $u_j$ its displacement with respect to the equilibrium position, $\alpha$ the harmonic coupling, and $\lambda$ the nonlinearity. Note that for later convenience we take a lattice size of $N+2$, but consider a constant chain length~\cite{barford} so $u_0=u_{N+1}=0$. We finally assume that the lattice motion only modulates the exciton hopping, so the exciton-lattice interaction is given by
\begin{equation} \label{ex_lat_classical}
H_{\text{in}}=\delta_J\sum_{j=1}^{N-1}\left(c_{j}^{\dagger}c_{j+1}+\text{h.c.}\right)\left(u_{j+1}-u_{j}\right),
\end{equation}
with $\delta_J$ the exciton-lattice coupling. In the absence of the quartic nonlinearity, this corresponds to the celebrated Su-Schrieffer-Heeger (SSH) model~\cite{su1988rmp}.

We now consider the quantized motion of the lattice, defining $\omega=\sqrt{2\alpha/m}$ and the creation ($a_j^{\dagger}$) and annihilation ($a_j$) phonon operators for site $j$ as~\cite{barford}
\begin{equation}
u_j=\sqrt{\frac{1}{2m\omega}}\left(a_j^{\dagger}+a_j\right),\quad p_j=i\sqrt{\frac{m\omega}{2}}\left(a_j^{\dagger}-a_j\right).
\end{equation}
Then we substitute $u_j$ and $p_j$ in Eqs.~\eqref{lat_classical} and~\eqref{ex_lat_classical}. The exciton-lattice interaction takes the simple form
\begin{equation} \label{int_quantum}
H_{\text{in}}=\Delta_J\sum_{j=1}^{N-1}\left(c_{j}^{\dagger}c_{j+1}+\text{h.c.}\right)\left(a_{j+1}+a_{j+1}^{\dagger}-a_j-a_j^{\dagger}\right)
\end{equation}
with $\Delta_J=\delta_J/\sqrt{2m\omega}$. Thus the exciton number is conserved, but the phonon number is not. The resulting lattice Hamiltonian is more complicated so we divide it into the linear and non-linear components, namely $H_{\text{lat}}=H_{\text{lat-lin}}+H_{\text{lat-nl}}$. For simplicity, to avoid unnecessarily complex Hamiltonian, here we only keep terms with equal number of phonon creation and annihilation operators, as these are the terms that remain under a rotating wave approximation; the others do not survive, as they oscillate more rapidly. Thus the linear terms of the lattice Hamiltonian are, up to a constant energy shift,~\footnote{Note that using the boundary conditions $u_0=u_{N+1}=0$ removes from the lattice Hamiltonian extra local terms on sites $j=0,N+1$, making it spatially homogeneous.}

\begin{figure}[t]
\begin{center}
\includegraphics[scale=0.3]{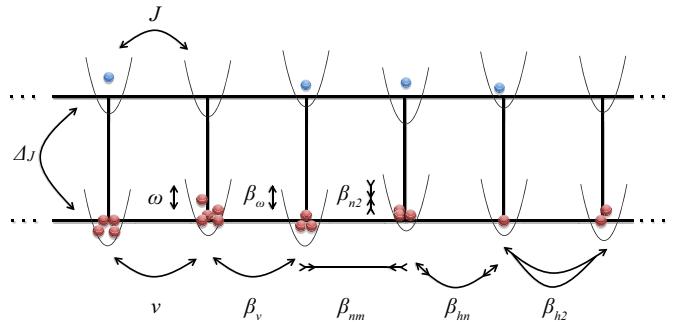}
\caption{\label{diagrama} Scheme of the system under study. The top chain represents the exciton lattice, where their hopping to nearest neighbors is described by the parameter $J$. This hopping is modulated by the coupling to phonon degrees of freedom, denoted by $\Delta_J$. The bottom chain, which correspond to the phonon lattice, is described by linear and non-linear components. There are three on-site terms, namely the linear on-site phonon energy $\omega$, its correction from the nonlinearity $\beta_{\omega}=9\beta$ and the non-linear on-site phonon interaction $\beta_{n2}=3\beta$. The remaining terms involve nearest-neighbor couplings. First, the direct hopping of phonons is given by the linear term $\nu$ in addition to the extra hopping $\beta_{\nu}=3\beta$ coming from the nonlinearity. The other non-linear terms are the following: $\beta_{nn}=6\beta$ is the nearest-neighbor density-density interaction, $\beta_{h2}=\frac{3}{2}\beta$ represents hopping of phonon pairs, and $\beta_{hn}=3\beta$ is the density-dependent hopping.}
\end{center}
\end{figure}

\begin{equation}
H_{\text{lat-lin}}=\omega\sum_{j=1}^{N} n_j-\nu\sum_{j=1}^{N-1}\left(a_j^{\dagger}a_{j+1}+\text{h.c.}\right),
\end{equation}
where the first term gives the on-site energies and the second one corresponds to phonon hopping with $\nu=\omega/4$. The nonlinearity of the lattice takes the form
\begin{subequations}  \label{non_linear_quantum}
\begin{align}
H_{\text{lat-nl}}=\beta&\Biggl[9\sum_{j=1}^{N}n_j-3\sum_{j=1}^{N-1}\left(a_j^{\dagger}a_{j+1}+\text{h.c.}\right)\label{non_linear_quantuma} \\
&+3\sum_{j=1}^{N}n_j^2+6\sum_{j=1}^{N-1}n_jn_{j+1}\label{non_linear_quantumb} \\
&+\frac{3}{2}\sum_{j=1}^{N-1}\left(\left(a_j^{\dagger}\right)^2a_{j+1}^2+\text{h.c.}\right) \label{non_linear_quantumc} \\
&+3\sum_{j=1}^{N-1}\left(a_j^{\dagger}a_{j+1}+\text{h.c.}\right)\left(n_j+n_{j+1}\right)\Biggr], \label{non_linear_quantumd}
\end{align}
\end{subequations}
with $\beta=\lambda/(2m\omega)^2$. This shows that several types of non-linear terms emerge. Equation~\eqref{non_linear_quantuma} corresponds to corrections to the linear terms, i.e. the phonon on-site energies and simple hopping which may be absorbed in the definitions of $\omega$ and $\nu$ of Equation (7). Equation~\eqref{non_linear_quantumb} contains local (first term) and nearest-neighbor (second term) density-density interactions. Equation~\eqref{non_linear_quantumc} represents hopping of boson pairs. Finally, Eq.~\eqref{non_linear_quantumd} is a density-dependent hopping, modulated by the populations of the sites involved in the process. We have represented in a pictorial way all the terms of the Hamiltonian in Fig.~\ref{diagrama}. With the different types of arrows we refer to: $<->$ hopping (horizontal) and on-site energies (vertical), $>-<$ repulsion, and $><-><$ terms with mixed characters. We also note that since the terms of Eq.~\eqref{non_linear_quantuma} involve two-phonon operators, while the rest involve four, we refer to them as low- and high-order non-linearities respectively. Throughout this paper we will analyze the impact of each one of these terms on the propagation of excitons. 

A few remarks are in order. First, we note that the vibrational degrees of freedom can also couple to the excitons through the modulation of on-site energies~\cite{Iubini2015njp}, affecting the system in a nontrivial way. However, for the sake of simplicity we have not considered this effect, and restrict solely to the off-diagonal SSH coupling of Eq.~\eqref{ex_lat_classical}. Second, we have neglected cubic non-linearities in the lattice Hamiltonian of Eq.~\eqref{lat_classical}, which are known to be unstable in the absence of even anharmonicities~\cite{ashcroft}. Since these terms result in an odd number of creation and annihilation phonon operators under quantization, they do not conserve phonon number in contrast to those of Eq.~\eqref{non_linear_quantum}, and do not survive under the considered rotating wave approximation. Going beyond this assumption might unravel novel transport enhancement or drag mechanisms, and is left as future research. Finally, note that in systems such as conjugated polymers, where the SSH model has been extensively applied, the exciton and phonon energy scales are well separated, the former being of a few eV while the latter is typically of $\hbar\omega\approx0.2$ eV ~\cite{tozer2012jpca,mannouch2018jcp,barford}. Thus a direct decay of excitons into phononic degrees of freedom can be neglected. 

\subsection{Analysis of exciton dynamics}

To analyze the dynamics in this model, we assume that excitonic excitations are created on the left boundary of the chain at time $t=0$, which immediately results in a distortion of the lattice. Thus a few phonons are also created in the sites near the excitons. We then simulate the propagation of the excitons and phonons across the lattice under different conditions, and identify those that favor exciton transport the most. Since the Hamiltonian is quite complex, we study the impact of different terms separately before considering all its terms simultaneously. We start by looking at the linear model only, focusing on the competition between phonon hopping and phonon-exciton interaction. Then we discuss the effect of each non-linear term (Eqs.~\eqref{non_linear_quantumb},~\eqref{non_linear_quantumc},~\eqref{non_linear_quantumd}). Finally we analyze the dynamics under the full Hamiltonian.

To calculate the time evolution of the initial state we use the time-dependent density-matrix renormalization group in the matrix product state formalism~\cite{vidal2004prl,schollwock2011ann}, implemented with the open-source Tensor Network Theory (TNT) library~\cite{tnt,tnt_review1}. This method fully accounts for the exciton-phonon correlations, which play a key role in the observed dynamics~\footnote{A mean-field decoupling between phonons and excitons leads to an effective interaction Hamiltonian $H_{\text{in}}$ which depends on expectation values $\langle a_j\rangle=0$ and Re $\langle c_j^{\dagger}c_{j+1}\rangle=0$, and thus does not capture how they affect each other. A similar problem arises when neglecting all spatial correlations within this approach, making it unsuitable for discussing the effects of interest.}. It also incorporates correlations between particles of the same species, which make the problem intractable by techniques based on Green's functions and the Landauer-B\"uttiker formalism, commonly used for quantum transport~\cite{datta}. Furthermore, it allows us to go beyond a polaron-like treatment for situations where the excitons and phonons propagate at very different velocities. For all the simulations we consider system sizes of $N=24$ and reach times $T=12$ (in units of inverse exciton hopping), so that if free, excitons propagating from the left reach the right boundary at the end~\footnote{The propagation speed of free excitons in a rigid lattice is $2aJ$, with $a$ the spacing between neighboring sites.}. We also restrict to the hard-core boson limit for excitons, but allow up to 3 phonons per site. This choice allows us to perform accurate simulations with a moderate computational effort. In fact we verified that a larger maximal phonon occupation leads to the same results. In addition, it remains valid up to high temperatures. Assuming independent harmonic oscillators with frequency $\hbar\omega\approx0.2$ eV, we estimate that the mean phonon population values in our simulations (from $2/24$ for initial states in of Sec.~\ref{nonlinear_section} to $\approx6/24$, see Fig.~\ref{num_phonon}) correspond to temperatures of $\sim10^3$ K.

\begin{figure}[t]
\begin{center}
\includegraphics[scale=0.33]{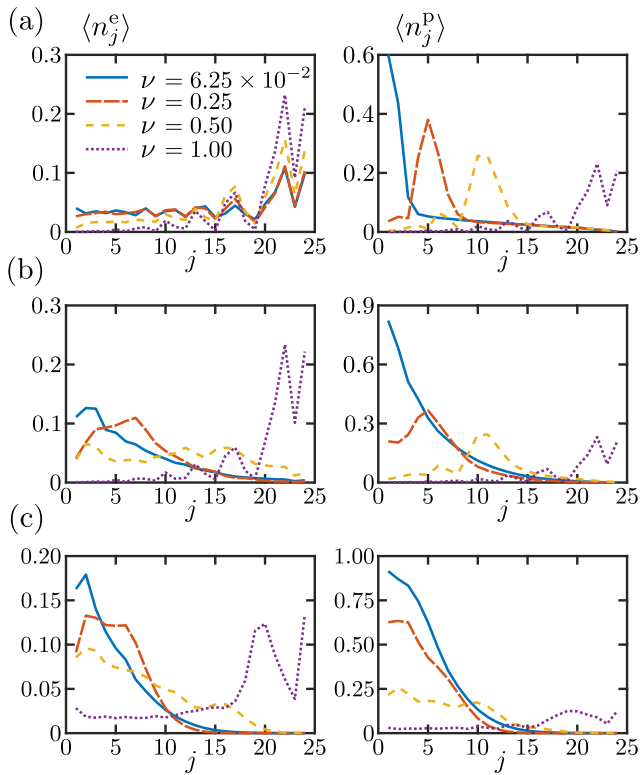}
\caption{\label{profiles_linear} Final population profiles of excitons (left) and phonons (right), for several values of $\nu$ and an initial single exciton-phonon excitation at the left boundary. (a) Weak exciton-phonon coupling $\Delta_J=0.1$. (b) Intermediate coupling $\Delta_J=0.25$. (c) Strong coupling $\Delta_J=0.8$.}
\end{center}
\end{figure}

We characterize the dynamics by obtaining the final spatial density profiles of excitons and phonons, i.e. the population per site in both chains at time $T=12$. In addition, to quantify how far the excitations propagate, we calculate the exciton and boson centers of mass (c.m.). They are defined by
\begin{equation}
R_{\text{c.m.}}^{\sigma}=\frac{\sum_{j=1}^{L}j\langle n_j^{\sigma}\rangle}{\sum_{j=1}^{L}\langle n_j^{\sigma}\rangle}
\end{equation}
with $\sigma=\text{e, p}$ for excitons and phonons respectively, and $\langle n_j^{\sigma}\rangle$ the corresponding densities at site $j$, i.e. $\langle n_j^{\text{e}}\rangle=c_j^{\dagger}c_j$ and $\langle n_j^{\text{p}}\rangle=a_j^{\dagger}a_j$.

\section{Linear dynamics: phonon hopping vs phonon-exciton coupling} \label{linear_section}

We begin our study focusing on the linear terms of the Hamiltonian, i.e. $\lambda=\beta=0$, and considering different values of phonon hopping $\nu$ and exciton-phonon coupling $\Delta_J$. To keep this analysis as simple as possible, we start by taking as initial state an exciton and a phonon at site $j=1$, with all the other sites being empty. This configuration already captures the basic physics of the competition between localizing and delocalizing effects, where strong exciton-phonon coupling and low phonon hopping largely suppress the exciton propagation.

\begin{figure}[t]
\begin{center}
\includegraphics[scale=0.65]{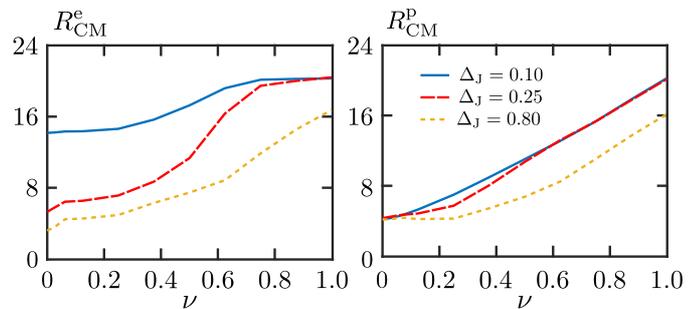}
\caption{\label{CM_linear} Final position of the c.m. for excitons (left) and phonons (right) as a function of $\nu$, for fixed values of $\Delta_J$ and an initial single exciton-phonon excitation on the left boundary of the lattice.}
\end{center}
\end{figure}

In Fig.~\ref{profiles_linear} we show the final population profiles of excitons and phonons for weak (upper panels), intermediate (middle panels) and strong (lower panels) exciton-phonon interaction $\Delta_J$. In addition, we plot in Fig.~\ref{CM_linear} the final location of the c.m. for the same $\Delta_J$ as a function of $\nu$. For weak coupling $\Delta_J=0.1$ the exciton and phonons move essentially independent of each other, as the wavefront of the former reaches the right boundary at the end of the simulations regardless of whether the phonons are fast ($\nu\approx1$) or remain almost localized ($\nu\ll1$); see Fig.~\ref{profiles_linear}(a). The coupling between both has a notable quantitative effect though, as the excitonic population that reaches the right edge is increased for faster phonons. This is also observed in the slow but significant increase of the excitonic c.m. position with $\nu$ depicted in Fig.~\ref{CM_linear}, which contrasts with the linear and faster increase of the phonon c.m.. When $\nu\rightarrow J=1$, the exciton dynamics feels a very weak drag from the phonon moving at approximately the same rate, and the position of its c.m. saturates with $\nu$.

Intermediate values of $\Delta_J$ have a larger impact on the dynamics, as illustrated in Fig.~\ref{profiles_linear}(b) for $\Delta_J=0.25$. Firstly, a significant exciton population remains close to the left boundary for low $\nu$. Also the wave packet of phonons spreads more smoothly across the lattice, as they experience a stronger drag from the faster excitons. This is consistent with the behavior of the c.m. depicted in Fig.~\ref{CM_linear}; while the phonon c.m. position slightly differs from that of weak coupling at low $\nu$, the exciton c.m. covers a much shorter distance, being delayed by the slower phonons. At high $\nu$, on the other hand, the dynamics is almost identical to that of weak coupling. 

If $\Delta_J$ is increased even more, up to the strongly-interacting regime, the exciton propagation is largely impeded if the phonons are not fast enough. As shown in Fig.~\ref{profiles_linear}(c), the exciton remains in the first half of the lattice along with the phonons up to $\nu=0.5$. Even for equal hopping in both lattices, there is a delay in the propagation of the excitonic and phononic c.m. with respect to weaker $\Delta_J$, as seen in Fig.~\ref{CM_linear}, and the left-most sites of the system remain significantly populated.

In summary, the impact of linear phonons on the exciton propagation unfolds as intuitively expected. If they are weakly coupled, the exciton propagates almost as a free particle, leaving the phonons behind if the latter have lower hopping. On the other hand, for strong interactions the excitons and phonons behave as a composite object (i.e. a polaron) with unified dynamics, where slow phonons can strongly slow down the exciton dynamics. We note that identical qualitative results were obtained for initial states involving more phonons and excitons (not shown), which are to be considered in the following Sections which discuss the impact of non-linear effects.  

\section{Effect of phonon non-linearities} \label{nonlinear_section}

\begin{figure}[t]
\begin{center}
\includegraphics[scale=0.65]{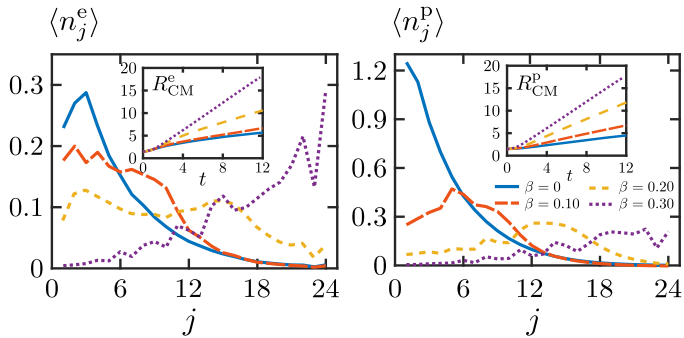}
\caption{\label{hopping_first_correction} Final exciton (left) and phonon (right) profiles for phonon dynamics with non-linear correction to linear hopping only, Eq.~\eqref{non_linear_quantuma}. The results are for $\nu=6.25\times10^{-2}$, $\Delta_J=0.25$ and several values of $\beta$. Insets: Corresponding c.m. position $R_{\text{c.m.}}^{\sigma}$ for $\sigma=\text{e, p}$ as a function of time $t$.}
\end{center}
\end{figure}

Having observed the main features of the competition between exciton and phonon hopping and their coupling, we proceed to analyze the impact of the vibrational non-linear terms in Eq.~\eqref{non_linear_quantum} on the exciton dynamics, with weak amplitude $\beta\ll1$. Given that these terms involve higher-order processes, it is necessary to consider more particles than in Sec.~\ref{linear_section}; otherwise the simulations underestimates the potential impact of the non-linearities. For this we consider an initial state of two excitons and two phonons, where there is one of each kind in each of the two left-most sites of the lattice. Also, since the strongest impact of phonons on excitons occurs when the former are slow, corresponding to a rigid lattice (low $\alpha$) or highly massive ions (large $m$), we consider $\nu\ll1$ from now on. Furthermore, we restrict our study to the intermediate-interacting case $\Delta_J=0.25$, which already features a strong impact of the lattice vibrations on the exciton dynamics. In addition, before considering the total nonlinearity of Eq.~\eqref{non_linear_quantum}, we unravel the effect of different terms separately. Our results show that some of them can largely enhance the exciton propagation, while others impede it.

\subsection{Correction to linear hopping} \label{first_nonlinear_correction}
We first discuss the impact of the non-linear correction to the linear phonon terms, namely Eq.~\eqref{non_linear_quantuma}. This includes a shift of the on-site energy, whose effect was seen to be negligible, and of the phonon hopping rate, which becomes $\nu+3\beta$. Thus a positive value of the non-linear interaction leads to a direct enhancement of the phonon propagation, naturally resulting in faster excitons than in the purely-linear case. This is observed in Fig.~\ref{hopping_first_correction}; the left panel shows how the exciton propagates further into the lattice as $\beta$ increases, going from a localized state in the linear case ($\beta=0$) due to the very slow phonons, to being largely delocalized for a weak nonlinearity $\beta=0.3$; the right panel shows the corresponding reach of the phonons.

These results illustrate the main observation of the present work. Namely, a weak nonlinearity of the underlying oscillating lattice of propagating particles can affect the dynamics of the latter significantly, even inducing a delocalized state which in the linear regime was highly localized. In addition, we found that for weak values of $\beta$, this non-linear correction to the phonon hopping has the strongest impact on the exciton propagation amongst the different non-linearities. Thus in the following, rather than analyzing each remaining term of Eq.~\eqref{non_linear_quantum} on its own, we explore the impact of each on the dynamics of the system already including the correction of Eq.~\eqref{non_linear_quantuma}. 

\subsection{Addition of density-density interactions} \label{sub_den_den}

\begin{figure}[t]
\begin{center}
\includegraphics[scale=0.62]{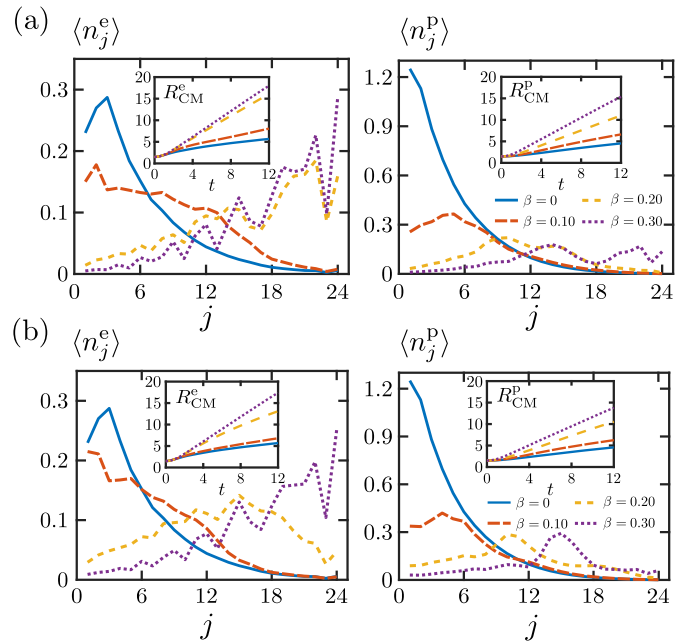}
\caption{\label{density_density} Final exciton (left) and phonon (right) profiles for phonon dynamics with non-linear density-density interactions, Eq.~\eqref{non_linear_quantumb}. The results are for $\nu=6.25\times10^{-2}$, $\Delta_J=0.25$ and several values of $\beta$. (a) On-site density-density interaction. (b) Nearest-neighbor density-density interaction. Insets: Corresponding c.m. position as a function of time.}
\end{center}
\end{figure}

Now we consider the impact of phonon density-density repulsion on the linear dynamics plus the non-linear hopping correction. As seen in Eq.~\eqref{non_linear_quantumb}, two contributions emerge, namely a local and a nearest-neighbor coupling, the former having half of the amplitude of the latter~\footnote{We have also performed calculations with each density-density interaction alone, and found that for them to have an important impact on the exciton dynamics (e.g. to induce delocalization), very large values of $\beta\geq1$ are required. Thus it is more illustrative for the weak $\beta$ regime to focus on their impact in the presence of the dominating non-linear term.}. The corresponding dynamics is shown in Figs.~\ref{density_density}(a) and~\ref{density_density}(b) respectively. It is intuitively expected that such interactions prevent phonon propagation, which is consistent with our results, as seen when comparing the right (main and inset) panels to those of Fig.~\ref{hopping_first_correction} for $\beta=0.2,0.3$. In particular, the on-site interactions reduce the phonon population of the right-most reached sites, while the nearest-neighbor couplings slow down the motion of the phonon wave packet as a whole. 

The qualitative impact of the two types of density-density couplings on the excitons is quite similar for most values of $\beta$, as seen in the left panels of Fig.~\ref{density_density}. However a notable difference occurs for intermediate values $\beta\approx0.2$, which is not too weak so the nonlinearity is barely present, yet not too strong so the correction to the hopping described in Sec.~\ref{first_nonlinear_correction} is dominant. There the on-site phonon repulsion delocalizes the excitons more, allowing them to reach the right edge of the system. Note that this delocalization is even stronger than that of the non-linear hopping correction alone (compare to the left main and inset panels of Fig.~\ref{hopping_first_correction}); so density-density phonon interactions also provide an enhancing mechanism for exciton transport.

\begin{figure}[t]
\begin{center}
\includegraphics[scale=0.62]{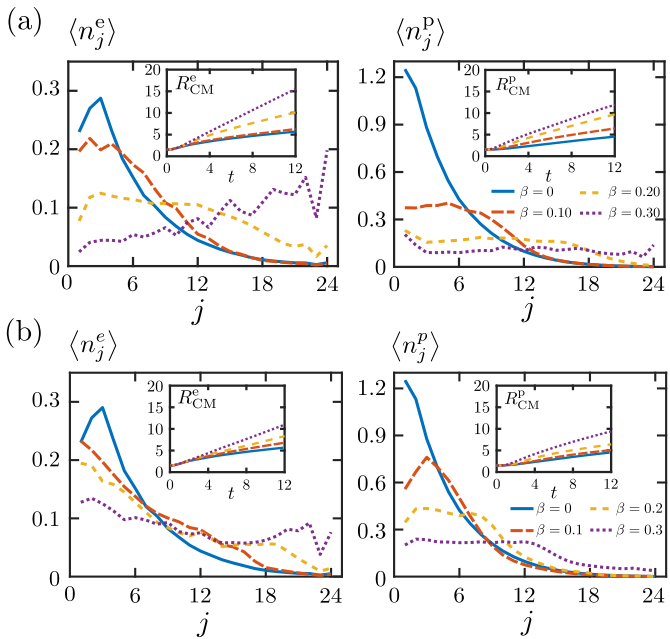}
\caption{\label{double_density_hopping} Final exciton (left) and phonon (right) profiles for higher-order non-linear phonon hopping processes, Eqs.~\eqref{non_linear_quantumc} and~\eqref{non_linear_quantumd}. The results are for $\nu=6.25\times10^{-2}$, $\Delta_J=0.25$ and several values of $\beta$. (a) Double phonon hopping. (b) Density-modulated phonon hopping. 
Insets: Corresponding c.m. position as a function of time.}
\end{center}
\end{figure}

This effect is better appreciated in Fig.~\ref{cm_vs_beta}(a), which depicts the c.m. position of excitons and phonons as a function of $\beta$ for different situations. There it is clearly seen that for most values of $\beta$, for which the phonons move more slowly than in the case of the hopping correction alone, the density-density repulsion enhances the exciton transport, with the maximal impact at $\beta=0.2$.   

The emergence of this phenomenon can be tracked to the total phonon number in the lattice, defined by
\begin{equation} \label{total_phonon}
n^{\text{p}}_{\text{T}}=\sum_{j=1}^{L}\langle n_j^{\text{p}}\rangle.
\end{equation}
As depicted in the left panel of Fig.~\ref{num_phonon}, the density-density interactions tend to reduce the phonon number in the system (due to the energy cost of having phonons close to each other) compared to the linear plus low-order non-linear correction. So there is a lower mass inducing a drag on the excitons, and the latter can move faster. This is specially stronger for the on-site repulsion, for which the total number of phonons is lower and which leads to the fastest dynamics for intermediate values of $\beta$.

\subsection{Addition of high-order non-linear hopping} \label{sub_hop}

\begin{figure}[t]
\begin{center}
\includegraphics[scale=0.62]{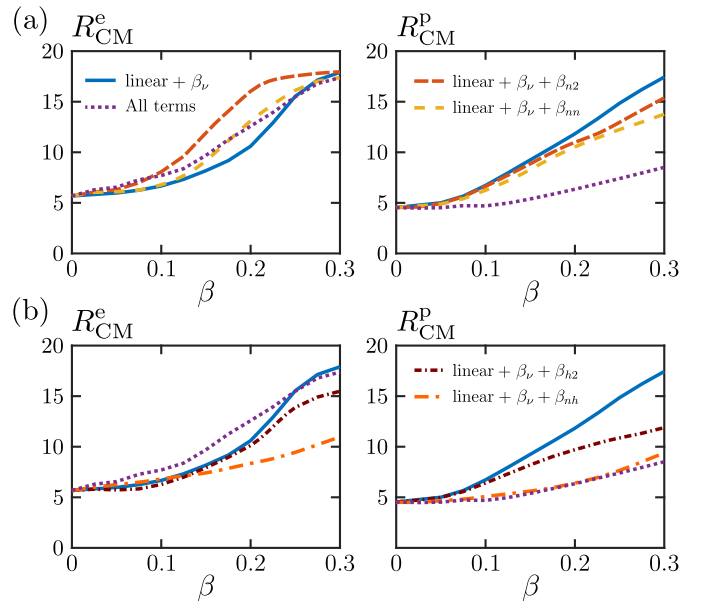}
\caption{\label{cm_vs_beta} Final position of the c.m. for excitons (left) and phonons (right) as a function of $\beta$, for the different non-linearities. All panels include the results for the linear plus the non-linear hopping correction of Eq.~\eqref{non_linear_quantuma}, and the total nonlinearity. The upper panels correspond to density-density non-linearities of Eq.~\eqref{non_linear_quantumb}, and the lower panels to the high-order hopping terms of Eqs.~\eqref{non_linear_quantumc} and~\eqref{non_linear_quantumd}.}
\end{center}
\end{figure}

We now discuss the impact of the higher-order non-linear hopping terms of Eqs.~\eqref{non_linear_quantumc} and~\eqref{non_linear_quantumd}, namely the boson-pair and density-mediated hopping processes. Since their sign is opposite to that of the linear hopping and its first non-linear correction, both are expected to slow down phonon propagation. As seen in the right panels of Fig.~\ref{double_density_hopping} and Fig.~\ref{cm_vs_beta}(b) for $\beta=0.2,0.3$ this is indeed the case, and this conduction impedance is even stronger than that caused by density-density interactions. In addition, the density-mediated hopping results in the slowest phonon dynamics amongst all the non-linearities, given that the modulation due to the densities of the sites involved in the process leads to a large effective (positive) amplitude. On the other hand, given the low amount of single-site phonons of the situation under study, the double-hopping mechanism is weaker.

\begin{figure}[t]
\begin{center}
\includegraphics[scale=0.63]{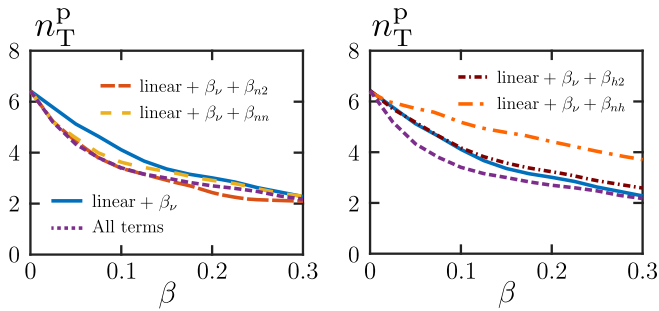}
\caption{\label{num_phonon} Final total phonon number as a function of $\beta$, for the different non-linearities. Both panels include the results for the linear plus the non-linear hopping correction of Eq.~\eqref{non_linear_quantuma}, and the total nonlinearity. The left panel correspond to density-density non-linearities of Eq.~\eqref{non_linear_quantumb}, and the right panel to the high-order hopping terms of Eqs.~\eqref{non_linear_quantumc} and~\eqref{non_linear_quantumd}.}
\end{center}
\end{figure}

The effect on exciton transport is opposite to that described in Sec.~\ref{sub_den_den}. Namely, as seen in the left panels of Fig.~\ref{double_density_hopping} and Fig.~\ref{cm_vs_beta}(b), excitons tend to become slower compared to the case of lower-order (linear plus non-linear) hopping when increasing both higher-order hopping mechanisms. This behavior can again be related to the impact of the Hamiltonian terms on the total phonon number defined in Eq.~\eqref{total_phonon}. As shown in the right panel of Fig.~\ref{num_phonon}, these non-linearities increase the number of phonons in the lattice, with the effect being stronger for the density-modulated hopping.

The overall picture uncovered by the previous results it thus that not only the speed of the phonons determine that of the excitons (e.g. the fastest phonons correspond to the linear plus low-order non-linear hopping). In addition, the number of extra phonons created by the particular nature of the nonlinearity, as well as the exciton-phonon coupling, induce a complex interplay which determines the resulting speed of exciton propagation.


\subsection{Dynamics under full nonlinearity}

\begin{figure}[t]
\begin{center}
\includegraphics[scale=0.63]{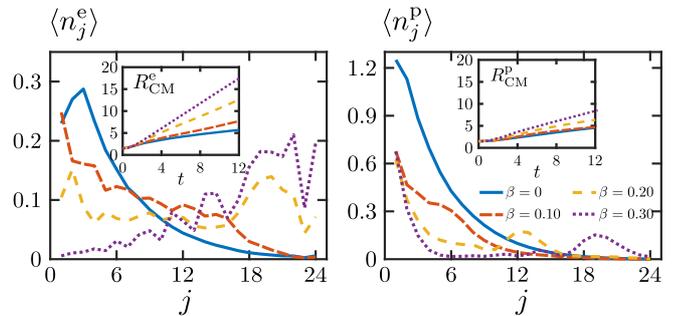}
\caption{\label{all_nonlinear} Final exciton (left) and phonon (right) profiles for phonon dynamics with the total nonlinearity, Eq.~\eqref{non_linear_quantum}. The results are for $\nu=6.25\times10^{-2}$, $\Delta_J=0.25$ and several values of $\beta$. Insets: Corresponding c.m. position as a function of time.}
\end{center}
\end{figure}

We finally discuss the system dynamics in the presence of the full non-linear Hamiltonian, Eq.~\eqref{non_linear_quantum}. The final population profiles are shown in Fig.~\ref{all_nonlinear}, which are qualitatively similar to those already discussed. The effect of the full phonon nonlinearity is thus to increase the exciton transport across the lattice, even if its amplitude $\beta$ is very weak. In addition we can determine the interplay between the different non-linear corrections, from the c.m. final positions as a function of $\beta$ depicted in Fig.~\ref{cm_vs_beta}.

For very weak non-linearities $\beta\lesssim0.1$, the density-density terms dominate, and the fastest exciton dynamics is found for the full combination of terms. For larger amplitudes $0.1\lesssim\beta\lesssim0.2$ the high-order non-linear hopping terms start having more weight in the overall dynamics, slowing down the exciton transport compared to the (fastest) on-site repulsion scenario. For even larger amplitudes $0.2\lesssim\beta\lesssim0.3$ the dynamics adopts an intermediate behavior, being slower that that of density-density interactions, but faster than the high-order non-linear hopping cases. Furthermore, we have observed that for $\beta=\mathcal{O}(1)$ all the curves of Fig.~\ref{cm_vs_beta} are almost indistinguishable, indicating that the $\beta_{\nu}$ correction of the linear hopping entirely dominates the dynamics. Since this is far from the regime of our interest (and indeed no convoluted physics is observed there), those results are not shown.  

Finally, in contrast to the exciton case, the full nonlinearity always leads to the slowest phonon dynamics, as seen in the right panel of Fig.~\ref{cm_vs_beta}. This occurs because, as discussed in Sec.~\eqref{sub_den_den} and Sec.~\eqref{sub_hop}, and in spite of the low total phonon number (see Fig.~\ref{num_phonon}), each high-order non-linear process decreases phonon propagation.

\section{Conclusions} \label{conclu}
In the present work we have analyzed the transport of excitons in the presence of an oscillating background lattice of ions, as described by an SSH model, with a weak quartic anharmonicity. By quantizing the vibrational degrees of freedom we have identified several underlying phonon-conserving processes, which have a different impact on the exciton dynamics. To unravel the effect of each term, we calculated the propagation of a few initially-localized particles in the system, and obtained the reach of each scenario by observing the final spatial population profiles and c.m. evolution. We performed our simulations using the time-dependent density matrix renormalization group, which incorporates the correlations between different components of the system efficiently and thus allows for the systematic and rigorous numerical study of a wide range of parameter regimes.

For the purely harmonic phonon dynamics we focused on the competition between phonon hopping and exciton-phonon coupling. We found that while fast phonons or weak couplings allow excitons to propagate across the lattice in a way similar to free particles, slow phonons and strong coupling strongly prevent exciton transport.   

For the anharmonic scenario we considered an intermediate exciton-phonon coupling with slow phonons (for linear slow dynamics) and observed the effect of different resulting non-linearities separately. First we identified that a non-linear (low-order) correction to the usual phonon hopping is the leading term, strongly enhancing exciton transport compared to the linear case. Then we observed that on top of it, density-density interactions can enhance the exciton transport even more for intermediate amplitudes of the nonlinearity. Subsequently we found that higher-order phonon hopping processes have the opposite effect on excitons, slowing them down compared to the linear plus low-order non-linear hopping. Such opposite effect arises from the higher amount of phonons created by the latter non-linearities, in contrast to the density-density interactions which penalize in energy their emergence. The overall effect of the non-linear lattice, resulting from the combination of the previous terms, is an enhancement of exciton transport in between both positive and negative effects.

The key point to emphasize from our results is that even a very weak lattice non-linear amplitude can have a large impact on the dynamics of the corresponding system, exemplified here in a strong increase of exciton propagation compared to the linear case. We expect that our results motivate the study of the impact on different weak non-linearities in strongly-correlated systems, such as a cubic anharmonicity, and of alternative particle-phonon coupling schemes (e.g. a local coupling as in Holstein models~\cite{chung2007prb,vidmar2011prb,golez2012prl,dorfner2015prb,brockt2015prb,hashimoto2017prb,brockt2017prb,tozer2012jpca,mannouch2018jcp}), to determine how they can affect their already very rich physics.

\begin{acknowledgments}
The authors thank Susana Huelga for useful discussions on the very early stages of the work. J.J.M.-A. thanks Luis Quiroga, Ferney Rodr\'iguez and Karina Guerrero for discussions. J.J.M.-A. thanks the Galileo Galilei Institute for Theoretical Physics for the hospitality and the INFN for partial support during the completion of this work. J.J.M.-A. also acknowledges financial support from Facultad de Ciencias at UniAndes-2015 project \textit{Quantum control of nonequilibrium hybrid systems-Part II}, and the support of Departamento Administrativo de Ciencia, Tecnolog\'ia e Innovaci\'on (COLCIENCIAS), through the project \textit{Producci\'on y Caracterizaci\'on de Nuevos Materiales Cu\'anticos de Baja Dimensionalidad: Criticalidad Cu\'antica y Transiciones de Fase Electr\'onicas} (Grant No. 120480863414). J.P is grateful for financial support from Ministerio de Ciencia, Innovaci\'on y Universidades (SPAIN), including FEDER funds: PGC2018-097328-B-I00 together with Fundación S\'eneca (Murcia, Spain) Projects No. 19882/GERM/15. M.B.P acknowledges support from the ERC Synergy grant BioQ.
\end{acknowledgments}

\bibliography{nonlinearity_bib}

\end{document}